# Nigeria's ICT and Economic Sustainability in the Digital Age

**Agbeyangi, A. O.**[*], **Makinde, A. S.**[**] **& Odun-Ayo, I. A.**[*]

[*]Department of Computer Sciences, Chrisland University, Abeokuta, Ogun State, Nigeria
[**]Department of Computer Science, Edo State University, Uzairue, Nigeria

Nigeria's remarkable information and communication technology (ICT) journey spans decades, playing a pivotal role in economic sustainability, especially as the nation celebrates its Republic at Sixty. This paper provides an overview of Nigeria's ICT journey, underscoring its central role in sustainable economic prosperity. We explore the potential of artificial intelligence, blockchain, and the Internet of Things (IoT), revealing the remarkable opportunities on the horizon. We stress the urgency of achieving digital inclusivity, bridging the urban-rural gap, and reducing the technological divide, all of which are critical as Nigeria marks its sixtieth year. We intend to prove the invaluable opportunities of ICT for policymakers, business leaders, and educational institutes as Nigeria looks towards enduring economic development in this digital age. Specifically, we envision a dynamic landscape where emerging technologies are set to redefine industries, supercharge economic growth, and enhance the quality of life for every Nigerian.

## 1. Introduction
### 1.1 Overview of Nigeria's remarkable ICT journey

Nigeria's ICT journey has been characterised by remarkable transformation, reflecting the nation's adaptability and commitment to the digital age (David & Omolola, 2015; Benedicta & Lacheheb, 2022). It started with issues like inadequate infrastructure and access restrictions, making telecommunication services a luxury that only a select few could afford. A significant turning point was the liberalisation of the telecommunications industry in 2001, which opened the door for private enterprises to join the market (Enahoro & Olawade, 2021). The liberalisation policy led to the entry of private companies such as MTN, Glo, Airtel, and Etisalat (now 9mobile) into the Nigerian telecommunications market. It has been described as one of the most successful economic reforms in Nigeria, as it has transformed the Nigerian telecommunications industry from a monopoly to a competitive market. This turning point specifically paves the way for improvement in the ICT revolution and has a major impact on the Nigerian ICT journey.

The development of the National ICT Policy unveiled in 2001 (Diso, 2005; Olatokun, 2006), provided a structured framework for ICT growth, emphasising local content development, innovation, and digital inclusivity. This policy has been essential in forming Nigeria's ICT strategy and guaranteeing that technology will always be at the centre of the country's economic growth. The emergence of technological firms like Andela[1], Flutterwave[2], and Paystack[3] shows how Nigerian startups have the ability to compete internationally while fostering innovation and economic expansion. These companies have not only attracted significant investment from both local and foreign investors, but they have also created thousands of high-skilled jobs for Nigerian youth. Furthermore, the government's commitment to promoting digital literacy and providing infrastructure support has paved the way for more startups to thrive in the country. With a strong ICT strategy in place, Nigeria is well-positioned to continue its technological advancements and drive sustainable economic development.

---
[1] https://andela.com/
[2] https://flutterwave.com/
[3] https://paystack.com/

The creation of jobs, innovation, and local content has generally been aggressively encouraged by government programmes, most notably those of the Nigerian Content Promotion and Development Board (NCPDB) and the National Information Technology Development Agency (NITDA). Though they still exist, issues like poor rural ICT infrastructure and cybersecurity worries are being addressed by different intervention programmes from the World Bank among others (Mbanaso et al., 2019). Another noteworthy accomplishment is Nigeria's entry into the global digital economy, as seen by the participation of Nigerian businesses in trade, outsourcing, and partnerships with leading global tech corporations.

Nigeria's ICT journey highlights the country's dedication to sustainable economic development in the digital era as it commemorates its 60th anniversary of independence. Nigeria looks forward to a time when new technologies will transform industries, quicken economic expansion, and improve everyone's quality of life.

## 1.2 Significance of Nigeria at Sixty in the context of ICT and economic sustainability

Nigeria's 60th Republic anniversary is a crucial moment in the country's efforts to utilise information and communication technology (ICT) for long-term economic stability. ICT has undergone significant advancements over the past sixty years and has emerged as a key catalyst for economic expansion in Nigeria. This shift has been facilitated by the liberalisation of the telecoms sector, government policies that prioritise the development of information and communication technology (ICT), and the emergence of tech firms.

Nigeria's celebration of its Sixtieth republic highlights the significant advancements made in narrowing the digital divide. Several projects, such as network infrastructure connectivity in both urban and suburban areas, which seek to broaden internet availability in rural regions, serve as a prime example of the country's dedication to ensuring that the advantages of information and communication technology (ICT) are accessible to all Nigerians. These initiatives have not only improved connectivity in remote areas but have also fostered economic growth and innovation. Furthermore, the government's commitment to fostering a conducive environment for tech startups has attracted foreign investments and propelled Nigeria's position as a leading hub for technology and entrepreneurship in Africa (Assay, 2021). Thus, bolstering economic prospects, particularly in marginalised areas.

Moreover, the sixtieth anniversary symbolises Nigeria's heightened incorporation into the worldwide digital economy. Nigerian enterprises are increasingly engaging in international trade, outsourcing, and forming partnerships with global IT industry heavyweights (Anwar & Graham, 2022). This integration not only amplifies the nation's impact in the global technology sector but also fosters commerce, ingenuity, and economic viability. Furthermore, the growth of entrepreneurship in Africa has also led to the emergence of innovative solutions to address pressing social and environmental challenges. Nigerian IT entrepreneurs are actively developing sustainable business models, such as renewable energy solutions and eco-friendly products, to contribute to the global fight against climate change. These initiatives not only promote environmental sustainability but also create job opportunities and drive economic growth in the country.

Nigeria foresees a future driven by digital advancements, where rising technologies revolutionise industries and expedite economic expansion. The Republic at Sixty emphasises the country's ability to recover from challenges, its ambitions, and its steadfast dedication to advancing towards a successful and technology-focused future, with information and communication technology (ICT) as a central component. This demonstrates Nigeria's unwavering commitment to persist in its impressive advancement in the field of information and communication technology (ICT), promoting sustainable economic growth for the foreseeable future. Nigeria recognises the transformative power of ICT and its potential to bridge the digital divide, create job opportunities, and enhance the delivery of essential services. The government has implemented various initiatives and policies to foster innovation and entrepreneurship in the ICT sector, attracting investments and nurturing a vibrant startup ecosystem. As a result, Nigeria has witnessed remarkable progress in expanding internet access, promoting digital literacy, and developing local tech talent. With a strong foundation in ICT, Nigeria is poised to become a leading player in the global digital economy, further solidifying its position as a key player in Africa's technological renaissance (Gómez et al., 2023).

## 2. Historical Perspective

The evolution of information and communication technology (ICT) in Nigeria has been a transformative journey that has opened new avenues for education, knowledge dissemination, economic growth, and social development. This journey, which began in earnest in the 1960s, has witnessed significant breakthroughs in the realms of computers, the internet, and the World Wide Web, fundamentally altering the country's landscape. As the most populous country in Africa, Nigeria's rapid urban growth over the past few decades has created an ideal environment for the proliferation of ICT, influencing various sectors such as education, health, agriculture, business, government, and transportation (Suleiman et al., 2020). The integration of ICT in these sectors has resulted in greater access to information and services, higher efficiency, and enhanced communication. Additionally, the government's investment in ICT infrastructure and policies has played a vital role in fostering digital inclusion and bridging the digital gap among different parts of the country.

### 2.1 Early Internet Adoption and Policy Initiatives

The path to widespread ICT adoption in Nigeria has not been without its challenges. In 2012, the country's internet usage stood at less than 16%, highlighting a significant digital divide (Agwu & Carter, 2014; Inegbedion, 2021). The average download speed at that time was a mere 1.38 Mbps, lagging far behind the 10.1 Mbps enjoyed in the United States. Despite these challenges, Nigeria's government recognised the transformative potential of ICT and took strategic steps to bridge the digital gap. In March 2001, the National Information Technology Policy was approved, laying the foundation for Nigeria's ICT development (Ibem et al., 2021). The establishment of the National Information Technology Development Agency further solidified Nigeria's position as a key player in the African ICT landscape. Over the last 15 years, Nigeria has leveraged ICT as a catalyst for sustainable development and international competitiveness.

The country outlined a visionary mission, aiming to become a globally competitive, knowledge-based society while integrating ICT into every aspect of its socio-economic development.

Nigeria's dedication to ICT development is apparent in its endeavours to narrow the digital gap and foster digital literacy among its populace. The government has enacted several measures, including the development of ICT training institutes and the deployment of broadband infrastructure, to provide broad access to technology and internet connectivity. These policies have not only given individuals and businesses more power, but they have also established Nigeria as a central location for innovation and technical progress in Africa (Gómez et al., 2023).

## 2.2 The Role of Mobile Communication

Mobile communication has played a pivotal role in Nigeria's ICT transformation. In 2013, a staggering 83% of the population, equating to 166.6 million people, were active mobile phone subscribers (Forenbacher et al., 2019). This rapid adoption of mobile technology showcased the immense potential ICT holds for transforming society. ICT not only facilitated communication but also had the power to reduce poverty rates and elevate the economic and social status of its citizens (Onyema, 2019). Moreover, the extensive utilisation of mobile phones in Nigeria has created prospects for financial inclusion and entrepreneurship. Mobile banking services have facilitated financial inclusion for those residing in distant regions, enabling them to engage in saving, investing, and actively participating in the formal economy. Also, the accessibility of mobile internet has facilitated the outreach of small enterprises to a broader clientele and the expansion of their activities, fostering economic development and the generation of employment opportunities (Jonathan & Escobar, 2010).

In furtherance, the enhanced connectivity offered by mobile phones has fostered the expansion of e-commerce in Nigeria. The advent of online markets and platforms has provided small-scale businesses with the opportunity to expand their customer base by selling their products and services to a wider audience, both locally and worldwide. This has not only augmented their earnings but has also generated job prospects for others who may now serve as delivery persons or offer auxiliary services for these internet enterprises. The emergence of e-commerce has also facilitated the development of new ideas and rivalry, resulting in enhanced goods and services across multiple sectors.

Mobile communication has also enhanced innovation and increased market accessibility for farmers in the agricultural sector (Ogunniyi & Ojebuyi, 2016; Oladele, 2011). This according to Ogunniyi & Ojebuyi (2016) contributes to increased income, reduced transaction and transportation costs, and increased farm productivity. Mobile applications and services based on Short Message Service (SMS) offer farmers up-to-date data on meteorological trends, market rates, and optimal agricultural methods. This enables farmers to make well-informed decisions, increase production, and gain access to wider markets for their agricultural products. Through the utilisation of up-to-date data on meteorological trends, farmers can enhance their ability to strategize and safeguard their crops against unfavourable weather circumstances. Also, knowing market prices enables farmers to engage in more favourable negotiations and optimise their earnings. Furthermore, the availability of agricultural best practices through mobile apps and SMS services empowers farmers to adopt innovative methods and improve their harvest quality, resulting in increased market competitiveness (Emeana et al., 2020). In general, mobile communication has completely transformed the agriculture sector by giving farmers more control and stimulating economic development in most rural areas.

Similarly, mobile communication has additionally enabled farmers to conveniently get financial services, including mobile banking and digital payment systems (Babcock, 2015). This enables farmers to securely oversee their financial matters, obtain loans, and conduct transactions without the necessity of real banking institutions. The convenience and efficiency of this assist in the reduction of financial obstacles and the promotion of financial inclusivity in rural regions. In addition, the advent of mobile communication has created fresh marketing prospects for farmers, enabling them to establish direct connections with consumers via Internet platforms and social media. This eliminates the need for intermediaries and enhances their profit margins.

In addition to its economic impact, mobile communication has revolutionised social and civic participation. Mobile-accessible social media platforms have emerged as potent instruments for social activity, dissemination of information, and involvement of citizens (Ajayi & Adesote, 2016; Fasae & Adegbilero-Iwari, 2016). These platforms are crucial in influencing public discussions and ensuring that those in power are held responsible. They have been used as a platform for underrepresented voices to be amplified and for grassroots movements to gather strength. Mobile communication has also enabled citizens to actively engage in influencing their communities by organising protests, and rallies, and raising awareness about significant social concerns. Furthermore, the immediate availability of news and information via mobile devices has facilitated citizens' ability to remain well-informed and actively involved in current events, thus promoting a society that is more knowledgeable and politically conscious.

The role of mobile communication in the influence of ICT in Nigeria is diverse and complex. It goes beyond simple connectivity to enable individuals to have economic, social, and sector-specific empowerment. Nigeria's ongoing ICT journey is expected to witness the transformative development of mobile communication, which would bring forth additional innovation, inclusion, and beneficial societal effects.

## 4. Current ICT Landscape

In recent years, Nigeria has witnessed significant growth in the ICT sector, driven by the increasing adoption of smartphones and affordable internet services, as well as the government's commitment to promoting digital innovation and entrepreneurship. Additionally, the rise of e-commerce and digital payment solutions has created new opportunities for businesses and consumers alike. As more Nigerians gain access to ICT services, there is a positive ripple effect on various sectors, including education, healthcare, and agriculture, leading to improved efficiency and productivity. Therefore, the ICT sector is expected to play a crucial role in Nigeria's economic development in the coming years. The government has recognised the potential of the ICT sector and has implemented policies and initiatives to support its growth. This includes the establishment of technology hubs and incubators, as well as providing training and funding opportunities for startups and entrepreneurs. With these efforts in place, Nigeria is poised to become a hub for technological innovation and digital transformation in Africa.

The approval of 5G technology in Nigeria in 2021 (*Nigeria - Information and Communications Technology*, 2023) and subsequent licensing to major operators like MTN, MAFAB

Communications, and Airtel signalled a new era of advanced connectivity in the country. This leap in technology promises to revolutionise various sectors and enhance Nigeria's position in the global ICT landscape. Nigeria has the potential for enhanced internet speeds and improved reliability with the adoption of 5G technology, facilitating seamless communication and data transmission. This development will open up new possibilities for creative solutions and enhanced services for sectors including healthcare, education, and e-commerce. Furthermore, by drawing more foreign investments and promoting technological developments domestically, Nigeria's standing in the global ICT market will be enhanced.

For the Nigerian workforce, this growth, which hinges on 5G technology, will also create new opportunities. With the increasing prevalence of high-speed internet, remote work and telecommuting will become increasingly viable, enabling professionals to work from their homes or any preferred location. This adaptability will not only enhance the equilibrium between work and personal life but also diminish traffic congestion and carbon emissions linked to commuting. Moreover, the heightened connectivity facilitated by 5G will allow professionals to collaborate and share knowledge more effectively, resulting in improved productivity and innovation across diverse sectors. For instance, a Nigerian software developer can be employed by a global technology corporation without the need to physically travel, resulting in savings of both time and money on travel costs. By collaborating with colleagues across several time zones, individuals can effortlessly contribute to projects, leading to expedited product development and enhanced productivity. Furthermore, by connecting freelancers and businesses with clients around the world, this improved connectivity will also give them more opportunities to reach a wider audience.

## 4.1 Government Initiatives and Partnerships

The Nigerian government's recognition of ICT as a catalyst for several sectors such as education, healthcare, agriculture, and manufacturing highlight its dedication to promoting ICT advancement. The government aggressively encourages collaborations between local ICT firms and international investors and assists in programs such as incubator hubs, youth innovation programmes, and science technology parks to foster an entrepreneurial ecosystem in the technology industry (Adhikari et al., 2021; Gómez et al., 2023). An example for instance is the 3 Million Technical Talent (3MTT) initiative coordinated by the National Information Technology Development Agency (NITDA) and the Federal Ministry of Communications, Innovation, & Digital Economy (Oloruntade, 2023). The programme is a component of the present government's Renewed Hope strategy and is specifically designed to promote Nigeria's technical expertise. It was reported that the initiative is a component of President Bola Ahmed Tinubu's plan to generate 2 million digital jobs by 2025. These programs have the objective of fostering the expansion of domestic ICT startups and enticing foreign direct investment in the industry. In addition, the government has enacted regulations aimed at enhancing internet connectivity and infrastructure, hence facilitating the accessibility of ICT services for both enterprises and individuals (Adeyemo, 2011; Awagu, 2020). The Nigerian government is strategically establishing itself as a centre for innovation and technological improvement by investing in ICT development. This initiative is expected to contribute significantly to economic growth and employment creation in the country.

The significance of encouraging cooperation between business and academia has also been acknowledged by the government. In order to accomplish this, they have formed alliances with universities and research organisations to facilitate the exchange of knowledge and the advancement of state-of-the-art technologies (Ogbaga et al., 2022). These collaborative efforts not only improve the calibre of ICT education and research in Nigeria but also provide students and researchers with the opportunity to interact with industry experts and acquire hands-on experience. The mutually beneficial partnership between academics and industry has the capacity to stimulate innovation and boost Nigeria's technology sector to unprecedented levels of success.

The government's efforts in information and communication technology (ICT) can also be observed through the intervention of state governments in Nigeria. Multiple states in Nigeria have enacted regulations and initiated ICT initiatives to entice investments and cultivate advantageous circumstances for technology-oriented organisations. Lagos State for example is a technology zone with the aim of fostering innovation, whilst many other states are actively pursuing many other ICT initiatives to stimulate growth driven by technology (Adhikari et al., 2021). These states are fostering the creation of startups, research and development centres, and other technology-related companies by establishing designated zones and implementing beneficial policies (Onaleye, 2023). The impact of this formation of startups and research centres can generate employment prospects and stimulate economic expansion, rendering Nigeria an appealing destination for both domestic and international investors. Thus, demonstrating the government's dedication to promoting innovation and establishing Nigeria as a prominent participant in the worldwide technology arena.

In addition, these partnerships have resulted in the establishment of innovation hubs and incubators, which offer a nurturing setting for aspiring entrepreneurs to convert their concepts into feasible products and services. The hubs provide essential resources including mentorship programs, financial access, and networking opportunities that are vital for the development and prosperity of startups. The existence of these hubs and incubators has engendered a feeling of community and friendship among people involved in the technology industry, promoting a culture of cooperation and sharing of information.

### 4.2 Private Sector Engagement and Foreign Investment

The private sector makes a significant global impact through its continuous technical collaboration and innovation. The participation of the private sector has been important in Nigeria's Information and Communication Technology (ICT) expansion story. Several multinational technology companies, such as Microsoft, Facebook, and others, have established innovation centres in Lagos to leverage the abundant technological expertise in Nigeria (Dano et al., 2020). These partnerships have not only resulted in the creation of job opportunities and the promotion of economic growth, but they have also played a crucial role in the overall progress of Nigeria's Information and Communication Technology (ICT) business. The presence of these international businesses has provided opportunities for indigenous entrepreneurs to gain expertise and forge relationships with influential individuals in the industry, fostering an atmosphere that promotes innovation and collaboration. Moreover, the surge in foreign investment has been crucial in reducing the disparity in digital accessibility and improving the availability of information and communication technology for the Nigerian population.

As international businesses continue to invest in Nigeria's ICT sector, the country is experiencing a significant transformation in its digital landscape. The influx of foreign investment has not only boosted the economy but has also played a pivotal role in bridging the digital divide and increasing access to technology for all Nigerians (Ahmed, 2007; Shirazi & Hajli, 2021). This has resulted in a more connected society and has opened up new avenues for education, entrepreneurship, and economic growth. Additionally, the presence of international companies has stimulated competition and innovation within the local market, driving advancements in technology and creating a thriving ecosystem for ICT development.

With the support and direction of international companies, these growing businesses can expand and diversify their operations, thereby creating more employment opportunities for the local labour market. Furthermore, the presence of international corporations has attracted highly skilled individuals from different regions of the continent, thereby enriching the talent pool and promoting the growth of the ICT sector in Lagos. This inflow of highly qualified workers has aided in knowledge sharing and teamwork, which has propelled technological developments in addition to contributing to the expansion of the ICT sector. The establishment of multinational corporations in Lagos and other regions of Nigeria has created chances for local enterprises to engage in collaboration and form partnerships with these global entities. This allows them to access global markets and extend their footprint beyond the domestic market. In general, the existence of international companies has had a significant influence on the ICT industry, driving its expansion and establishing the city as a centre for innovation and technological advancement.

### 4.3 National Digital Economy Policy and Strategy

In 2020, Nigeria not only embraced the digital era but also underwent a complete transformation. The National Digital Economy Policy and Strategy (NDEPS) (Lamid et al., 2021) served not only as a policy document but also as a definitive statement of purpose. Nigeria declared its aspiration to not only participate in the digital realm but to excel and become a prominent figure, a pioneer, and a champion in the worldwide digital ecosystem. Nigeria embarked on the adoption of NDEPS with the aim of revolutionising its economy and establishing itself as a prominent participant in the worldwide digital market. This strategy has the overarching goal of enhancing digital infrastructure, advancing digital literacy, and cultivating innovation and entrepreneurship within the nation. Nigeria's adoption of the digital revolution demonstrates its commitment to utilising technology for economic advancement and societal progress. This brave move showcases Nigeria's digital capabilities and ambitions to the global community.

The substantial enhancement in Nigeria's digital infrastructure due to the deployment of the NDEPS is vital. The government made substantial investments in the expansion of broadband infrastructure, thereby guaranteeing access to high-speed internet even in distant regions. This action not only closed the gap in access to digital technology within the country but also established Nigeria as a centre for digital creativity and business. By implementing a strong digital framework, businesses of all sizes were able to utilise technology to expand into untapped markets, optimise their processes, and stimulate economic development.

NDEPS is fostering a dynamic network of startups, incubators, and accelerators, where ambitious concepts are translated into tangible outcomes, and young Nigerians are actively constructing the digital future, rather than merely engaging with it (Lamid et al., 2021).

Ultimately, the Emerging Technologies pillar focuses on anticipating future developments. NDEPS is geared towards the adoption of artificial intelligence, blockchain, and the metaverse, guaranteeing that Nigeria remains at the forefront of the digital revolution. The NDEPS provides a clear and strategic plan, rather than a final goal or endpoint. Nigeria is steadfastly and resolutely progressing on an ongoing path. Through steadfast dedication and cooperation, this digital economy will not only be constructed but will also serve as a symbol of groundbreaking progress and success for all Nigerians, showcasing their united advancement in the digital realm.

**IV. Emerging Technologies in Nigeria**
**A. Exploration of AI, blockchain, and IoT in the Nigerian context**
Nigeria, being the most populous nation in Africa, is actively adopting modern technology to promote innovation and stimulate economic growth. The focal point lies on the revolutionary technologies of Artificial Intelligence (AI), the Internet of Things (IoT), and Blockchain. Artificial Intelligence (AI), the science of creating intelligent machines, is transforming everything from medical diagnosis to gaming. It gains knowledge from data, whether labelled for supervised learning, unlabeled for unsupervised discovery, or reinforced by rewards. The wide range of capabilities offered by AI, including machine learning, deep learning, natural language processing (Eludiora et al., 2015; Agbeyangi et al., 2021), and computer vision, is revolutionising our world. The future promises even more extraordinary possibilities. Artificial intelligence (AI) is being widely adopted in various industries in Nigeria such as healthcare, finance, agriculture, and education (Arakpogun et al., 2021; Mhlanga & Aldieri, 2021). It is being used to improve disease diagnosis, provide personalised financial services, optimise agricultural yields, and create customised educational content. AI algorithms in the healthcare industry facilitate accurate diagnoses, while in finance, they enhance security and customer service.

Blockchain is essentially a network of interlinked blocks, where each block contains a specific piece of data and is securely tied to the preceding block (Yli-Huumo et al., 2016). This network establishes a digital fabric of consistent clarity, a characteristic that enables a wide range of opportunities in various sectors. Examine the current financial environment: a significant number of individuals are still unable to participate in conventional banking systems, preventing them from pursuing their aspirations due to limited accessibility. Blockchain-based solutions such as mobile wallets and microloans can create opportunities for financial inclusion, enabling those without access to traditional banking services and promoting a fairer economy.

However, the potential of blockchain extends beyond just financial transactions. The intricate intricacy of supply chains, where things swiftly move in and out of obscurity, can be controlled through their openness. Each stage, starting from the farm and ending at the shelf, is recorded in a secure ledger, reducing the occurrence of fraudulent activities and enhancing effectiveness. This has the potential to revolutionise industries such as agriculture and logistics (Kamilaris et al., 2019). Envision a scenario where farmers are remunerated justly for their agricultural produce, and consumers have unwavering trust in the excellence of the products they purchase.

The healthcare industry has the potential to tremendously profit from the implementation of blockchain technology (Makinde et al., 2023). The healthcare sector encounters a multitude of obstacles, such as data confidentiality, compatibility, and counterfeit medications. These difficulties can be efficiently resolved by utilising blockchain technology. Blockchain technology can guarantee the secure and confidential exchange of patient data between healthcare professionals, enhancing the overall standard of treatment and facilitating improved patient results. Moreover, blockchain can be employed to monitor and authenticate the genuineness of pharmaceutical products, hence mitigating the threat of counterfeit medications in the market. Succinctly, the implementation of blockchain technology in the healthcare sector has the capacity to completely transform the industry and lead to substantial improvements in patient care and safety.

Moreover, land in Nigeria, a highly valuable asset, can become a battleground for conflicting claims of ownership. Blockchain, on the other hand, provides a peaceful sanctuary. An impregnable and lucid land register constructed upon its basis has the capability to eliminate the ambiguity and intricacy of land administration while ensuring the protection of legitimate possession. This might signify the preservation of family legacies and the resolution of disputes in a just and efficient manner.

The Internet of Things (IoT) is a broad network of networked devices that gather, share, and use data in real-time. It's an information-driven silent symphony that changes everything from our daily lives to our workplaces, turning innocuous items into sentient beings (Balaji et al., 2019). The potential of this digital revolution is vast. The integration of sensors in agriculture optimises irrigation techniques to enhance crop yields. Remote patient monitoring revolutionises healthcare by enabling doctors to seamlessly monitor vital signs in real-time. The movement of traffic resembles a coordinated ballet performance, facilitated by the instantaneous communication between connected vehicles. However, this intricate dance of data is not without its difficulties. Security requires attention and focus to ensure that confidential information is protected in IoT devices. Preserving privacy is essential to avoid an accidental jumble of personal information. The infrastructure, which comprises the digital cables of this interconnected world, requires meticulous building.

However, Nigeria, with its dynamic technology ecosystem and strong government backing, is well-positioned to confidently embrace this digital future. Entrepreneurs are creating novel Internet of Things (IoT) solutions, such as solar-powered irrigation systems and interactive classrooms equipped with smart screens (Atayero et al., 2016). Nigeria is prepared to participate in the global Internet of Things (IoT) movement and contribute its own distinct contribution to the digital landscape. The future is filled with insightful murmurs, and Nigeria is prepared to attentively listen and acquire knowledge. IoT is transforming transportation, energy, and urban development by enhancing traffic management, monitoring energy consumption, and enabling the construction of intelligent cities.

These technologies have significant potential for fostering Nigeria's economic growth. Artificial Intelligence (AI), for example, provides enhanced healthcare, banking, agriculture, and education, thereby augmenting the quality of life for Nigerians. The Internet of Things (IoT) enables enhanced efficiency in transportation and energy management, while Blockchain technology guarantees both transparency and security in the realms of finance and supply chains. These technologies are anticipated to persist in their incorporation into Nigeria's socioeconomic landscape, fostering innovation and prosperity. As these technologies progress and become increasingly available, they have the capacity to transform multiple sectors in Nigeria.

The emergence of advanced technologies, including artificial intelligence (AI), robotics, the Internet of Things (IoT), blockchain, cybersecurity, and big data, has the potential to make products and services more accessible to persons who have difficulty accessing them. Individuals living in rural locations can get significant advantages from utilising Internet banking, whereas individuals with limited mobility can profit from online purchasing and voting. Online education and remote work can provide viable options for parents who need flexibility in managing their houses or individuals who have caregiving obligations. These technology innovations can enable distant relationships with loved ones, improve our general standard of living, and empower us to make more ecologically aware choices. In addition, digital communication tools such as video conferencing and social media enable individuals with physical limitations to establish connections with others and engage in social activities that may otherwise provide challenges. In addition, technology has facilitated the accessibility of information for those with visual impairments through the utilisation of screen readers and speech recognition software. In general, these innovations have the capacity to dismantle obstacles and establish a more comprehensive society for individuals, irrespective of their accessibility limitations.

Moreover, upcoming technologies are positioned to transform our work life. The growing automation of monotonous and tedious chores will enhance our productivity and efficiency in the workplace, allowing us additional time and capability to participate in high-value, performance-oriented activities. Technological solutions that facilitate remote work offer the opportunity to pursue highly desired occupations without the need to move or face interruptions while travelling internationally. While some manual occupations may become outdated, the job market will provide a wide range of fresh prospects, characterised and propelled by rising technologies. Individuals will need to consistently acquire new skills and adjust in order to remain pertinent in the changing environment. Consequently, although emerging technologies may result in the elimination of certain occupations, they will ultimately generate a plethora of captivating opportunities for individuals who are open to embracing change and harnessing the potential of technology.

### III. The Role of ICT in Economic Sustainability
Information and Communication Technology (ICT) is essential for stimulating economic growth and development, and this is clearly demonstrated in the Nigerian environment. Although the formal sector of the Nigerian economy has reliable data to substantiate its activities, the informal sector's impact on the nation's GDP has frequently been disregarded. The swift integration of digital technology, along with the growing number of telecom customers and internet users, clearly demonstrates the significance of ICT in promoting economic sustainability, particularly in the informal sector. With the increasing availability of digital technology, individuals in the informal sector can enhance their enterprises and extend their client reach. This not only enhances their efficiency and financial gain but also adds to the overall expansion of the Nigerian economy. Moreover, the utilisation of Information and Communication Technology (ICT) in the informal sector facilitates enhanced clarity and effectiveness in commercial activities, hence enticing increased investments and fostering economic stability. Hence, it is imperative to acknowledge and utilise the capabilities of ICT in the informal sector to ensure long-lasting economic progress in Nigeria.

In addition, the use of digital technology into the disconnected economy has the potential to generate employment opportunities and mitigate poverty. With the increasing adoption of ICT tools by individuals and enterprises in the informal sector, there is a growing need for proficient professionals who can effectively operate and maintain these technologies. This fosters job prospects for the indigenous populace, particularly for the younger generation who frequently bear a disproportionate burden of unemployment. By granting individuals access to digital technology, the informal sector has the potential to stimulate economic empowerment and social inclusion, thereby elevating individuals from poverty and enhancing their overall well-being. Furthermore, the utilisation of ICT can augment the overall efficacy and competitiveness of the informal sector. By utilising digital tools, firms may optimise their operations, mechanise processes, and minimise expenses. This enables them to provide products and services at more favourable rates, thereby attracting a larger consumer base and expanding their market share. Moreover, the utilisation of digital platforms and online marketplaces facilitates the connection between enterprises operating in the informal sector and a broader customer base, both within the country and across borders. This presents novel prospects for advancement and enlargement, enabling organisations to increase in size and achieve more sustainability in the long term.

Nigeria, with its substantial population, has continuously observed ICT make up around 10 percent of the country's Gross Domestic Product (GDP). According to Prof Umaru Garba Danbatta, the Executive Vice Chairman of the Nigerian Communications Commission (NCC), the telecommunications industry's investment in the country's Gross Domestic Product (GDP) amounted to $70 billion in 2017 and has been steadily increasing ever since. The contribution of ICT to Nigeria's GDP increased to 10.5 percent in 2018, demonstrating consistent development. The consistent expansion of ICT's contribution to Nigeria's GDP signifies the growing significance of the industry in the nation's economy. The increasing adoption of technology and digitization by organisations has led to a growing need for ICT services. Not only does this enhance economic growth, but it also generates employment prospects and stimulates creativity within the nation. Nigeria possesses the capacity to leverage ICT and stimulate enhanced economic growth in the future, provided that appropriate infrastructure and regulations are established. For instance, the emergence of e-commerce platforms in Nigeria has greatly bolstered the impact of information and communication technology (ICT) on the country's gross domestic product (GDP). The emergence of online retail giants such as Jumia and Konga has fundamentally transformed the shopping habits of Nigerians, resulting in a notable surge in consumption and economic activity. Moreover, the proliferation of mobile banking services has enhanced the accessibility and convenience of financial transactions for Nigerians, hence fueling the advancement of Information and Communication Technology (ICT) in the economy.

ICT has not only bolstered the economy but has also played a pivotal role in developing camaraderie and solidarity among different tribes, facilitating participation in the informal sector despite intermittent crises. Similar to how football may overcome obstacles, ICT has also brought together the informal sector, promoting commerce, confidence, and commitment among diverse communities. By utilising digital platforms and social media, individuals belonging to diverse tribes in Nigeria have successfully established connections and engaged in the exchange of ideas, effectively dismantling enduring cultural boundaries. Not only has this enhanced relationships, but it has also stimulated collaboration and creativity within the informal sector. Furthermore, the convenience of communication enabled by ICT has encouraged rapid mobilisation and response in times of crisis, ensuring the informal sector's resilience and activity in the face of adversity. In summary, the impact of Information and Communication Technology (ICT) on the cohesion and involvement of the informal sector has played a crucial role in Nigeria's socio-economic progress.

The enhanced interconnectivity and dissemination of ideas among tribes in Nigeria have resulted in a more profound comprehension and admiration of one another's cultural heritage. This has not only cultivated a feeling of cohesion but has also facilitated cooperation and ingenuity within the informal sector. ICT has facilitated the dismantling of cultural boundaries, fostering an inclusive atmosphere that enables individuals from diverse tribes to collaborate, exchange their specialised knowledge, and collaboratively address shared obstacles. This has led to the emergence of novel products, services, and business models that have had a beneficial influence on the socio-economic framework of Nigeria. Moreover, the capacity to communicate rapidly and effectively via ICT has demonstrated its essentiality during moments of emergency. The informal sector has demonstrated rapid mobilisation, effective coordination, and timely assistance to impacted populations, whether in the face of natural disasters or health emergencies. The durability and adaptability of ICT have further emphasised the significance of ICT in sustaining and enhancing the informal sector's participation in Nigeria's socio-economic development.

**B. Case studies of successful ICT-driven economic initiatives**
ICT's impact on trade is one that stands out. ICT applications such as Facebook and WhatsApp are now being utilised as a novel means for conducting business transactions inside the informal sector. These platforms have not only offered financial prospects for several Nigerians but also acted as online stores for vendors who may lack the means to lease physical premises. The advent of digital trading has elicited positive responses from numerous entrepreneurs and households (Ukwuoma, 2019). The digitalization of trade in Nigeria has not only provided happiness to business owners and households, but it has also stimulated economic expansion and progress. For instance, the implementation of the mobile payment platform M-Pesa in Kenya has significantly transformed the nation's economy by enabling individuals to conveniently transfer and receive funds, settle debts, and get financial services using their mobile devices. This has facilitated the integration of the hitherto unbanked population into the financial system and has made a substantial contribution to the country's economic advancement. Estonia is another notable case study, renowned for its effective adoption of e-governance and digital services. This has resulted in a flourishing digital economy and the attraction of foreign investments. These examples illustrate the vast potential of ICT-driven initiatives in promoting economic progress and generating possibilities for both individuals and enterprises.

Nigeria has numerous instances of ICT efforts that serve as case studies and success stories. An illustrative instance of this phenomenon is the rapid expansion of the technology sector in Lagos, which has enticed substantial international capital inflows. Another notable example is the entertainment sectors in Nigeria, which have undergone a significant growth in recent years as a result of the rising use of ICT. These chances have allowed young people in Nigeria to utilise their skills and knowledge to develop creative solutions that effectively tackle local problems. Consequently, the Nigerian entertainment business has emerged as a substantial contributor to the nation's economy. Examining the prospects of content creators and social media influencers, it is clear that Nigeria's entertainment industry will persistently flourish and grow in the coming years. Mr. Macaroni, Broda Shaggi, Mark Angel, and Taaooma have achieved widespread fame and have become well-known figures in the entertainment sector. Their humorous skits and relatable material have struck a chord with millions of Nigerians, resulting in a devoted fan base and profitable sponsorship agreements. Moreover, the emergence of streaming platforms such as Netflix and YouTube has furnished Nigerian films and music with a worldwide stage, hence amplifying the industry's expansion. Given the government's ongoing allocation of funds towards infrastructure development and the growing availability of smartphones and Internet connectivity, Nigeria's entertainment sector is well-positioned to achieve even greater success in the years to come.

Moreover, the rise of social media has had a substantial impact on the triumph of Nigeria's entertainment sector. Social media platforms like Instagram, Twitter, and TikTok have facilitated direct interaction between artists, content creators, and their fans, enabling them to share their work and establish their individual identities. This direct engagement has not only facilitated artists in acquiring greater recognition but has also enabled them to obtain crucial comments and ideas from their audience. Consequently, Nigerian performers have successfully customised their content to better align with the interests and tastes of their fans, thereby contributing to the industry's expansion.

The case of payment platforms in Nigeria through ICT advances is of great significance, as it has brought about a revolutionary solution to the numerous issues encountered in the banking sector. Technology advancements like Flutterwave, Opay, and palmpay have greatly aided in the evolution of the banking industry by making transactions more convenient and available to a wider audience. These payment channels have not only facilitated the transfer of funds for individuals, but they have also created favourable conditions for the growth and success of small enterprises and entrepreneurs. The emergence of e-commerce in Nigeria has facilitated effortless online transactions, hence enhancing the expansion of the digital economy. In addition, the presence of reliable and effective payment options has enticed international investors, resulting in a significant increase in financial resources flowing into the Nigerian banking industry.

In general, the expansion of ICT advances in payment systems has been crucial in the development and advancement of Nigeria's financial industry. An illustrative instance occurred when a cash shortage occurred prior to the Nigeria 2023 election, causing difficulties for numerous Nigerians in accessing their monies and carrying out financial operations. The payment platforms had a substantial impact during this period by offering an alternate method for individuals to conveniently access their funds and conduct financial transactions.

The infusion of funds into the banking sector as a result of developments in information and communication technology (ICT) has played a significant role in the progression and growth of Nigeria's financial industry. The rise in capital has facilitated the growth and upgrading of banking infrastructure, resulting in enhanced services and more financial inclusivity. Furthermore, the progress in information and communication technology (ICT) has also enabled the expansion of digital banking, simplifying the distant access to financial services for both consumers and businesses. Undoubtedly, ICT plays a crucial role in Nigeria's financial industry, significantly contributing to innovation and propelling the sector's progress.

The current developments in education brought about by ICT have also been amazing and have significantly increased the access to learning possibilities for students of all ages. These technological breakthroughs have facilitated the creation of internet-based educational platforms and the utilisation of digital resources, empowering students to conveniently access educational materials and engage in virtual classrooms regardless of their geographical location within the country. Students in remote places, who previously had limited access to quality education, have greatly benefited from this. The incorporation of ICT in education has moreover facilitated interactive and captivating learning encounters, cultivating critical thinking and problem-solving aptitudes among students. In Nigeria, ICT has had a crucial impact on transforming the education industry, opening up new opportunities for learning and acquiring knowledge. An illustrative instance of how the incorporation of ICT has transformed education in Nigeria is the adoption of online learning platforms. These platforms provide students a diverse array of instructional resources, such as interactive videos, e-books, and practice exercises. Students have the opportunity to learn at their own speed and participate in self-guided learning, enabling them to have personalised educational experiences. Virtual classrooms further improve learning results by allowing students to participate in peer-to-peer collaborative activities and receive immediate feedback from teachers.

ICT has achieved widespread success across various areas of Nigeria's economy. Some further instances are the transportation and health care sectors. ICT has significantly transformed the delivery of medical services in the healthcare sector. Telemedicine enables patients residing in remote regions to establish virtual connections with doctors and specialists via video calls, therefore surmounting geographical obstacles and enhancing accessibility to high-quality healthcare. In addition, electronic medical records have optimised the organisation of patient information, facilitating healthcare personnel in monitoring and scrutinising data for enhanced diagnosis and treatment.

ICT has significantly enhanced productivity and safety in the transportation industry through the utilisation of GPS-enabled devices and applications. These sophisticated gadgets and applications, like as Google Maps, assist in monitoring and controlling the movement of vehicles, offering guidance to alleviate traffic congestion and mitigate accidents. Moreover, the advent of internet platforms catering to taxi and ride-sharing services like Uber, Bolt, and others has significantly enhanced the convenience and accessibility of commuting, hence amplifying the efficacy of transportation networks.

**VI. ICT Challenges and Future Directions**
Nigeria's ICT sector still faces difficulties despite its impressive progress. The issues encompass insufficient progress in infrastructure development, insufficient financial resources allocated to research and development, and a scarcity of proficient ICT experts. Moreover, the disparity in technology usage and accessibility between urban and rural areas continues to be a substantial obstacle to the adoption and utilisation of information and communication technology (ICT).

Nevertheless, the Nigerian government has acknowledged these difficulties and is actively pursuing measures to tackle them through the implementation of laws that encourage investment in information and communication technology (ICT) infrastructure and education. Additionally, the government is promoting collaborations with private sector organisations. The future trajectory of Nigeria's ICT sector is auspicious as these endeavours persist in narrowing the divide and unleashing the complete potential of ICT for economic expansion.

The growth of the sector has also been affected by delays in permit processing, various taxation levels, regulatory difficulties, and damage to existing infrastructure. Notwithstanding these obstacles, the Nigerian government has been diligently striving to surmount them. The government has implemented efficient permit processing systems, simplified taxation policies, and enacted regulatory reforms to foster a conducive environment for the growth of the information and communication technology (ICT) sector. In addition, endeavours are underway to mend and enhance the current infrastructure in order to guarantee a dependable and effective ICT network throughout the nation. These activities are anticipated to have a favourable effect on the growth of the sector and facilitate the development of a prosperous digital economy in Nigeria.

In addition, the government has prioritised the promotion of digital literacy and skills enhancement to ensure that the workforce possesses the requisite knowledge and ability to effectively harness the capabilities of information and communication technology. Diverse training programmes and initiatives have been put into effect to narrow the gap in digital access and provide persons with the necessary skills to engage in the digital economy. The government's objective in investing in education and training is to improve the skills and capacities of its residents, as well as to attract foreign investment and generate employment possibilities in the ICT sector. Adopting this comprehensive approach to nurturing the ICT sector is essential for Nigeria to establish itself as a prominent figure in technology and innovation within the region. The potential for the ICT sector to contribute to economic growth and social development is enormous, given the expanding population and rising internet penetration. The government's dedication to tackling the difficulties and establishing a conducive atmosphere is evidence of its acknowledgment of the transformative potential of ICT in Nigeria.

Nigeria's ICT sector is poised for a promising future, thanks to ongoing initiatives such as the GLO 2 underwater cable, the 2Africa project, and Google's Equiano cable. These projects are expected to enhance connectivity and lower internet costs. These projects will enhance internet connectivity for individuals and create favourable conditions for businesses to prosper in the digital economy. Moreover, the government's emphasis on cultivating domestic technology expertise and promoting innovation through endeavours such as tech hubs and incubators reinforces Nigeria's status as a centre for technological progress in the area. Investors have numerous opportunities in the fields of cloud services, digital financial services, fibre optics, broadband, satellite communication, and the provision of technology products. Allocating funds to these industries will not only bolster Nigeria's digital economy but also entice international investors seeking to capitalise on the nation's immense opportunities. Nigeria's substantial and young population creates a significant need for cutting-edge technical solutions, rendering it a highly appealing market for tech industry enterprises. Furthermore, the government's dedication to establishing a conducive business climate and enacting laws that promote innovation will enhance the prosperity of these investments.

Moreover, Nigeria's robust entrepreneurial culture and flourishing startup ecosystem create an ideal environment for the growth and success of technology enterprises. The nation has already witnessed the emergence of prosperous indigenous businesses, such as Jumia, Paystack, and Andela, that have garnered global acclaim and secured substantial investments. This dynamic environment not only promotes innovation but also generates employment opportunities for the country's expanding population. Moreover, Nigeria's advantageous geographical position in Africa establishes it as a pivotal entry point to the continent's burgeoning digital market, enabling enterprises to reach a wider range of customers and extend their operations on a regional level. Investing in Nigeria's digital economy is a prudent corporate decision that also aids the country's overall economic progress.

## X. Conclusion

Nigeria's ICT journey, which was characterised by important turning points and reforms to technology policy, is now seen as evidence of the country's adaptation and resilience in the digital era. The deregulation of the telecoms industry and the subsequent formulation of the National ICT Policy laid a vital groundwork for Nigeria's technological advancement. The Republic at Sixty represents a period of transformation, highlighting the crucial role of ICT in promoting economic sustainability by addressing infrastructure obstacles and cultivating a competitive market. Nigeria's ascent as a global contender in the digital economy is reinforced by the emergence of thriving businesses and the government's dedication to ensuring digital inclusion.

Looking into the future, Nigeria's Republic at Sixty represents not only a significant historical achievement but also a model for a prosperous and technologically advanced future. The country's dedication to inclusive growth is seen in its commitment to bridging the digital gap through infrastructure development and connectivity initiatives. Nigeria's thriving digital ecosystem, characterised by indigenous entrepreneurs and global partnerships, establishes the country as a significant participant in the worldwide technology industry. As the celebration progresses, it is clear that there is a strong dedication to maintaining this progress, with an emphasis on promoting creativity, generating employment prospects, and tackling environmental issues through technology-based solutions.

Nigeria anticipates a future in which emerging technologies fundamentally reshape industries and make a substantial contribution to economic growth. The Republic at Sixty symbolises the nation's lasting determination, ability to bounce back from difficulties, and unwavering commitment to progress in the digital age. Nigeria is well-positioned to manage the difficulties of the global digital economy and establish itself as a leader in Africa's technological rebirth, thanks to the significant role played by information and communication technology. As the country enters the next stage of its digital development, the determination to achieve a successful and technology-driven future remains strong, guaranteeing that the revolutionary impact of Information and Communication Technology (ICT) will continue to change Nigeria's economy for many years.